\documentclass[%
preprint,
showpacs,preprintnumbers,
 amsmath,amssymb,
 aps,
pra,
floatfix,
]{revtex4-1}

\usepackage{graphicx}
\usepackage{dcolumn}
\usepackage{bm}

\begin{document}

\title{Transmission of Megawatt Relativistic Electron Beams\\ Through Millimeter  Apertures}

\author{R. Alarcon,$^4$  S. Balascuta,$^4$ S.V. Benson,$^2$ W. Bertozzi,$^1$ J.R. Boyce,$^2$, R. Cowan,$^1$ D. Douglas,$^2$  P. Evtushenko,$^2$ 
 P. Fisher,$^1$ E. Ihloff,$^1$ N.Kalantarians,$^3$ A. Kelleher,$^1$ R. Legg,$^2$ R.G. Milner,$^1$  G.R. Neil,$^2$  L. Ou,$^1$ B. Schmookler,$^1$  C. Tennant,$^2$ C. Tschal\"ar,$^1$  G.P. Williams,$^2$ and S. Zhang$^2$}

\affiliation{$^1$Laboratory for Nuclear Science, Massachussetts Institute of Technology, Cambridge, MA 02139}
\affiliation{$^2$ Free Electron Laser Group, Thomas Jefferson National Accelerator Facility, Newport News, VA 23606}
\affiliation{$^3$Department of Physics, Hampton University, Hampton, VA 23668}
\affiliation{$^4$ Department of Physics, Arizona State University, Glendale, AZ 85306}

\date{\today}

\begin{abstract}
High power, relativistic electron beams from energy recovery linacs have great potential to realize new experimental paradigms  for pioneering innovation in fundamental and applied research.  A major design consideration for this new generation of experimental capabilities is the understanding of the halo associated with these bright, intense beams.  In this Letter, we report on measurements performed using the 100~MeV, 430~kWatt CW electron beam from the energy recovery linac at the Jefferson Laboratory's Free Electron Laser facility as it traversed a set of small apertures in a 127~mm long aluminum block. Thermal measurements of the block together with neutron measurements near the beam-target interaction point yielded a consistent understanding of the beam losses.  These were determined to be 3 ppm through a 2~mm diameter aperture and were maintained during a 7~hour continuous run.
\end{abstract}

\pacs{41.60.Cr, 41.75.Fr, 41.85.-p}
\maketitle

The Energy Recovery Linear Accelerator (ERL)~\cite{Tig1965}  offers significant advantages for delivery of relativistic electron beams for research: low emittance, small spatial size, high brilliance, and high power.  Successful operation of an ERL based on superconducting RF~\cite{Neil2000} has sparked interest over the last decade in groups worldwide to seriously pursue ERL technology for new applications.  These include light sources~\cite{Cornell2012}, electron-ion colliders~\cite{eRHIC2010}, and the delivery of electron beams to cool high energy hadron beams~\cite{Lit2007}.  

At present, there is significant interest~\cite{PEB2013} in the possibility to utilize Megawatt power electron beams from ERLs at energies of order 100 MeV on windowless gas targets to carry out new types of experiments to explore subatomic matter at low momentum transfers.  The measurements reported here are motivated by the optimal design of the proposed DarkLight experiment~\cite{DLprop2012} at the Jefferson Lab Free Electron Laser (FEL), which will search for a new boson beyond the Standard Model with a mass in the range 10 to 100 MeV/c$^2$.  This type of experiment demands a thick, windowless, gas target which can be achieved by flowing gas through low conductance apertures with transverse dimensions of order millimeters.  The target gas leaking through these tubes would be pumped away in stages to maintain vacuum in the beam pipes.  To minimize the size and cost of these vacuum pumps and to maximize the gas target density, the tube diameters need to be minimized. At the same time, beam losses in traversing the tubes need to be kept extremely small to minimize background.   The experiment is feasible only if the power losses of the Megawatt electron beam as it traverses these narrow conductances is tolerable and the resulting background rates are manageable.

\begin{figure}[htbp]
\centering\includegraphics[width=0.5\textwidth,height=0.15\textheight]{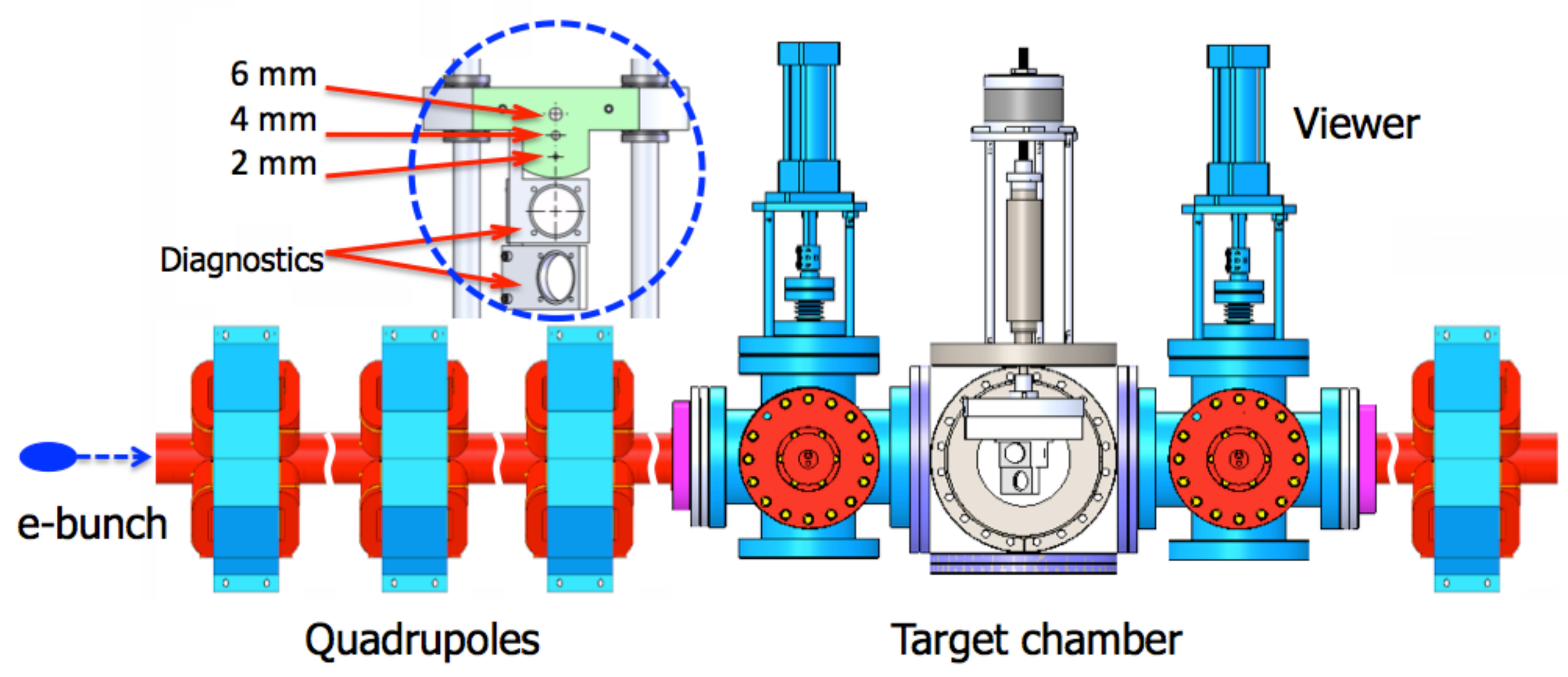}
\caption{Schematic layout of the experiment at the FEL beam facility at the Jefferson Laboratory.  For ease of illustration, the drawing is not to scale.}
\label{DL_test_layout}
\end{figure}

The measurements reported here were carried out using the 100~MeV electron beam from the ERL at the 3F region on the IR beamline at the FEL facility at the Jefferson Laboratory with the apparatus shown schematically in Fig.~\ref{DL_test_layout}.  At a modified section of the FEL beam (see Fig. 1) between two quadrupole triplets, a remotely controllable aperture block made of aluminum containing three apertures of 2, 4, and 6 mm diameter and 127 mm length was mounted in the beam pipe.  The block also carried an Optical Transition Radiation (OTR) crystal and a YAG crystal viewed by TV cameras to measure beam profiles and beam halo at the position of the aperture block. Any of these apertures or profile monitors could be placed on the beam axis by remote control. The temperature of the aperture block was monitored by a resistance temperature detector. The block temperature, beam current, repetition rate, and bunch charge were recorded.  Neutron and photon background monitors were placed near the aperture block and around the beam lines and the linac and their readings were continuously recorded.

The beam requirements for the transmission test were to achieve the maximum average beam current with small momentum spread and r.m.s. beam radius $\ll 1$ mm as well as minimal beam halo outside a 1~mm radius at the test aperture.  The details of the accelerator and beamline configuration for this test are discussed in \cite{Doug2012}. The small momentum spread was provided by ``cross-phased'' linac operation with the beam accelerated on the rising part of the RF wave in the first and third (low-gradient) accelerator module and on the falling part of the RF wave in the second (high-gradient) module. The phase-energy correlations so induced then cancel each another resulting in a small relative momentum spread of order 0.2\% f.w.h.m.

Minimal size of the core beam at the aperture was achieved by two alternate-gradient quadrupole triplets up- and down-stream of the aperture, producing a ``mini-beta'' region of $\beta\approx$ 0.2 m and an r.m.s. beam radius of $\approx$ 100 $\mu$m at the aperture. Additional quadrupoles and a full complement of beam monitors near the test region allowed beam phase advance adjustment, beam matching, and halo control without excessive betatron mismatch.

The moderate bunch charge of 60 pC minimized emittance and halo at the source. The small momentum spread alleviated the impact of dispersion errors, suppressed momentum tails, and mitigated effects of increased chromaticity of the ``mini-beta'' section. The longitudinal matching process (cross-phasing) allowed a long bunch reducing resistive-wall (wakefield) effects in the aperture.

\begin{figure}[htbp]
\centering\includegraphics[width=0.4\textwidth,height=0.3\textheight]{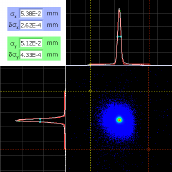}
\caption{Transverse beam profile distribution measured using the OTR at the aperture position.  The  frame enclosed between the cursors is 2 mm x 2 mm.
The best Gaussian fit (red lines) to the horizontal ($x$) and vertical ($y$) projections (white lines) of the transverse profile are also shown.  The $\delta \sigma_{x,y}$ are the uncertainties in the fit parameters given by the non-linear least squares fit of the Gaussian model to the data.}
\label{DL_transp4}
\end{figure}

Starting with a low-power beam, longitudinal match, lattice dispersion, and betatron match was established and subsequently repeated  after each beam power increase. The ``mini-beta'' section was tuned by first centering and logging the beam orbit through the 6~mm and subsequently the 2~mm apertures. After inserting the beam viewers, the beam spot at the aperture position was then minimized, as shown in Fig.~\ref{DL_transp4}

The beam down-stream of the test region was then retuned for low loss and zero beam break-up (BBU).  Subsequently, several combinations of bunch charges and repetition rates were tested for minimal background and aperture block heating in transmission through the 4 and 2~mm aperture. Finally, with fixed 60-pC bunch charge, beam transmission through the 2~mm aperture was optimized for increasing steps of beam power, fine-tuning the beam after each step, until it reached its full power of about 450 kW (4.5 mA, 100 MeV).

Halo losses were monitored by ion chambers and photomultiplier tubes and were kept minimal by beam tuning. Fine adjustment of beam steering and focussing near the aperture region kept the aperture block temperature rise and neutron and photon backgrounds minimal. Although a round beam spot of $50~\mu$m radius was achieved, $100~\mu$m spot radii at the aperture were typical.

A novel feature was BBU caused by small energy shifts. Because of the small momentum spread of 0.2\%, the usually observed Landau damping of BBU by tune spread from much larger momentum spreads was absent. Small energy shifts coupled to the chromaticity shifted the vertical phase advance which led to the onset of BBU. Stability was maintained by monitoring and minimizing the vertical beam size.  In the final 7-hour transmission run through the 2~mm aperture, the machine instabilities (particularly BBU) were controlled by stabilizing the energy at injection and in the recirculator.

\begin{table}[htbp]
\centering
\begin{tabular}{|c|c|r|r|c|c|c|r|c|}
\hline\hline
Run & apert. & duration  & $\Delta T$ & $T_{ave}$ & $P_B$ & $P_b$ & charge & $I_{ave}$ \\
\hline
        &    mm        &   min          & $^o$C                 &     $^o$C              &   W   &   MW   &    C   &   mA\\
\hline\hline
1 & 6 & 22  & 0.2  & 31.4  & 0.32  & 0.350  & 4.6 & 3.50 \\
2 & 4 & 30  & 0.7  & 31.6  & 0.56  & 0.372  & 6.7 & 3.72 \\
3 & 2 & 124  & 10.5  & 42.6  & 1.95  & 0.425  & 31.6 & 4.25 \\
4 & 2  & 413  & 9.1  & 43.8  & 1.09  & 0.422  & 121.0 & 4.22 \\
\hline\hline
\end{tabular}
\caption{Transmission Results.}
\label{Table1}
\end{table}

\begin{figure}[htpb]
\centering\includegraphics[width=0.5\textwidth,height=0.3\textheight]{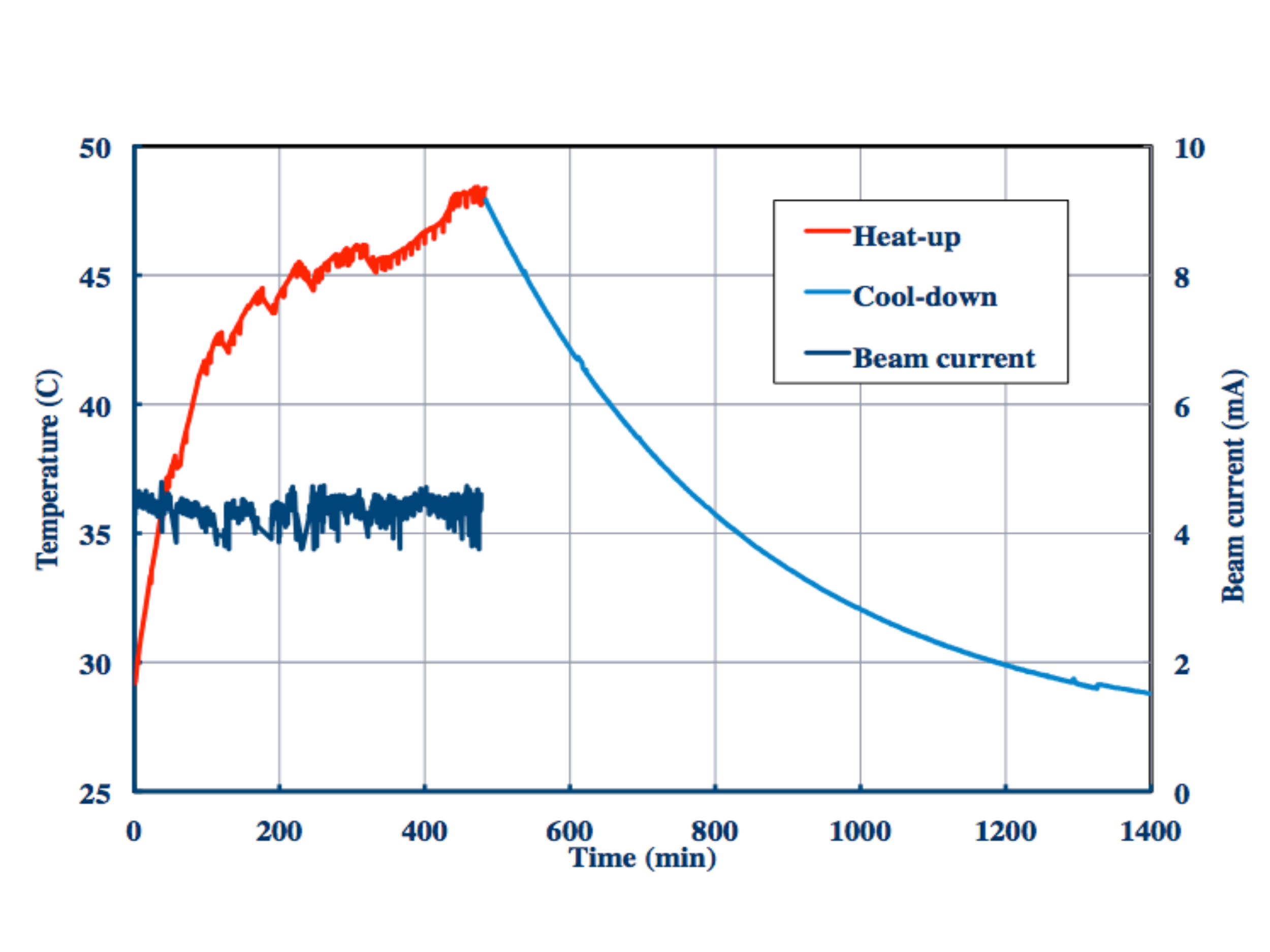}
\caption{Block temperature vs. time showing beam heating (in the presence of beam) and cooling (in the absence of beam).}
\label{DL_transp9}
\end{figure}

After optimizing the beam transmission through the apertures, four runs were taken, as summarized in Table~\ref{Table1}. There were two runs, Nos. 1 and 2, of 22 minutes and 30 minutes duration through the 6~mm and 4~mm apertures and two runs, Nos. 3 and 4, of 124 minutes and 413 minutes duration through the 2~mm aperture. Fig.~\ref{DL_transp9} shows the beam current and temperature rise for run No. 4 and the temperature fall without beam as functions of time. 

The power $P_B$ deposited by the beam in the aperture block is
\begin{equation}
P_B=c_pm(\frac{dT}{dt})_{block}+P_C
\end{equation}
\noindent
where $c_p$m = 917~Joule/$^o$C and $(\frac{dT}{dt})_{block}$ are the heat capacity and the rate of change of block temperature, respectively. $P_C$ is the power lost from the block through heat conduction and radiation to the beam pipe.

From the cooling data (Fig.~\ref{DL_transp9}), we found an exponential cooling time $\tau$ = 375 min and a base temperature of $T_o=27.2^oC$. The integrated beam energy deposited in the block during a run from time $t_1$ to $t_2$ is then
\begin{eqnarray}
E_B=\int_{t_1}^{t_2}dt\cdot P_B = c_pm[\Delta T+(T_{ave}-T_0)\frac{\Delta t}{\tau}]
\end{eqnarray}
\noindent
where $\Delta T = T(t_2)-T(t_1)$ is the temperature rise during the run time $\Delta t=t_2-t_1$ and $T_{ave}$ is the average temperature during the run. The temperature rise $\Delta T$ and the average temperature $T_{ave}$, deposited power $P_B$, and beam power $P_b$ as well as the total charge and average beam current for each run are summarized in Table~\ref{Table1}.

The power of the beam halo intercepted by the aperture block was only partly deposited in the block. A substantial part of the electromagnetic shower generated by the intercepted electrons escaped through the back and the sides of the block. Using the FLUKA  code~\cite{FLUKA}, a simulation showed that about 50\% of the energy of the halo electrons entering the block near the 2~mm aperture is deposited in the block. The remaining shower escaped through the sides and the downstream face of the block. It propagated down the beam pipe and was eventually absorbed in the pipe and surrounding beam line components.

\begin{figure}[htbp]
\centering\includegraphics[width=0.5\textwidth,height=0.35\textheight]{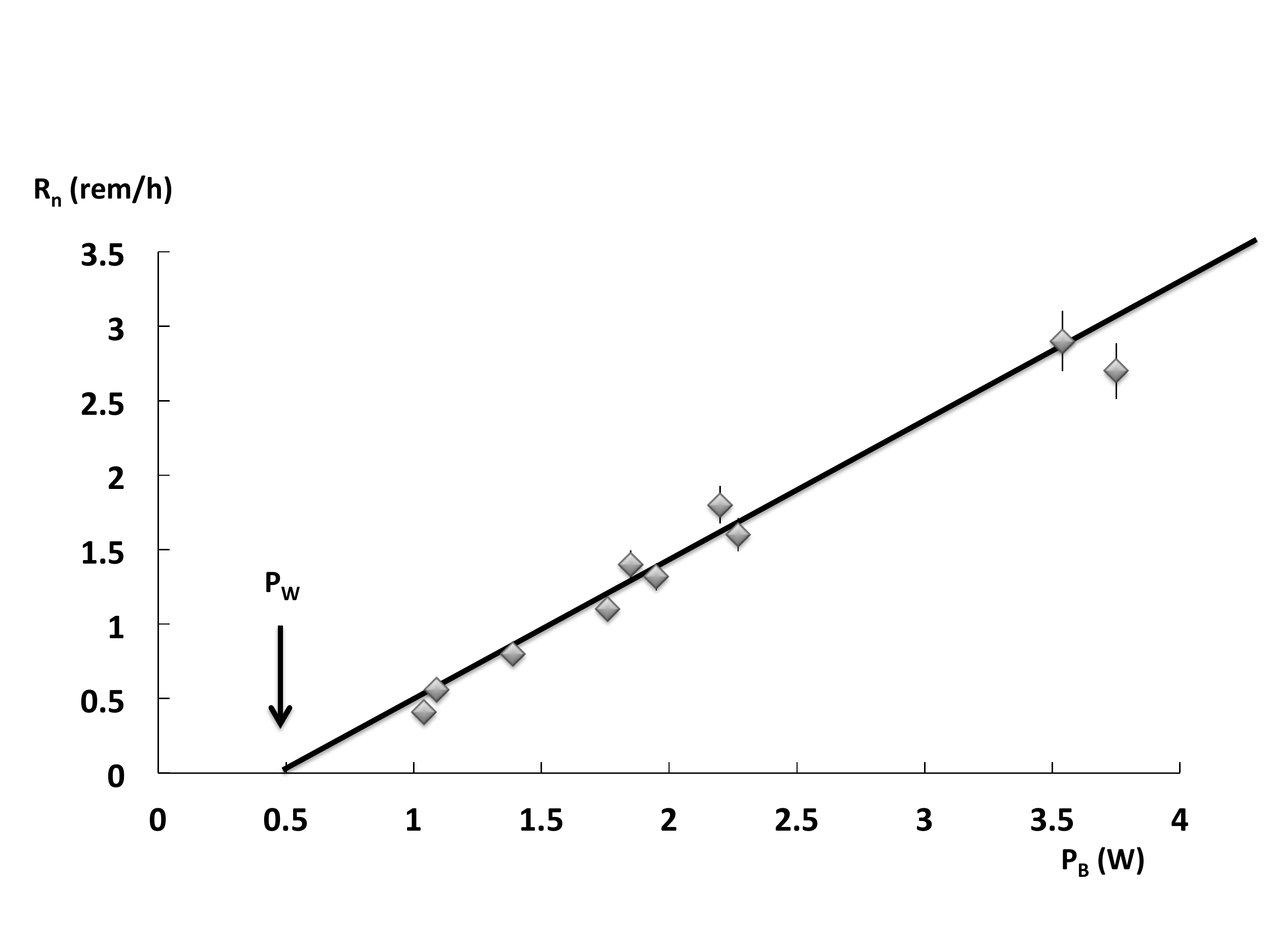}
\caption{The measured neutron dose rates vs. block power $P_B$.}
\label{Nratevspower}
\end{figure}

The neutron fluxes from the aperture block and the surrounding beam pipe were measured by a Canberra NP100B neutron rem-counter positioned 1.9 m downstream of the aperture block and $24^o$ to the left of the beam axis.  To relate the measured neutron fluxes to the power $P_B$ deposited in the block for a range of beam halo conditions, a selection of short sections of runs 3 and 4 where the neutron flux was reasonably stable were evaluated individually. The neutron dose rates $R_n$ plotted versus the block power $P_B$ are shown in Fig. ~\ref{Nratevspower}.

From the plot in Fig.~\ref{Nratevspower}, a linear relation was deduced of the form
\begin{equation}
R_n=0.9[\frac{\rm rem}{(Wh)}]\cdot(P_B-P_W)
\end{equation}
\noindent
where the straight-line intercept $P_W\approx$ 0.5 W was interpreted as the power deposited by the wake fields of the beam. Since the r.m.s. width of the beam at the aperture was less than about 0.1 mm or ten times smaller than the 2-mm aperture, it is reasonable to assume that the wake fields are largely governed by the bunch charge and time structure of the beam which was kept fixed throughout all four transmission runs.

As FLUKA simulations have shown that only about 50\% of the total power of the intercepted beam halo is deposited in the block, this power, $P_H$, is about twice the difference between $P_B$ and $P_W$ such that
\begin{equation}
R_n\approx 0.45[\frac{\rm rem}{(Wh)}]\cdot P_H
\end{equation}

\begin{figure}[htbp]
\centering\includegraphics[width=0.5\textwidth,height=0.2\textheight]{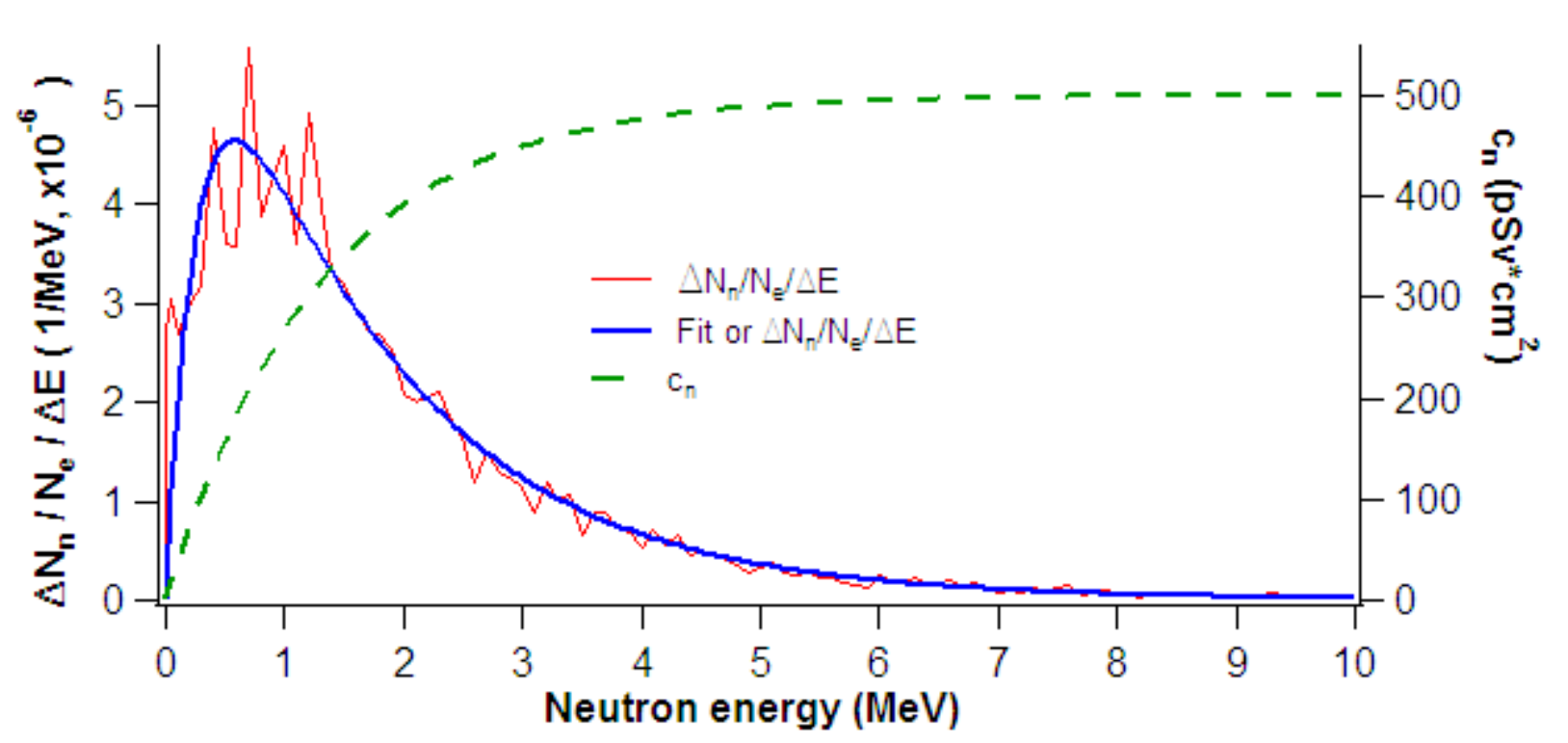}
\caption{ The neutron energy spectrum from the MCNP~\cite{MCNP} simulation (left axis) and the effective dose conversion factor $c_n$~\cite{Pell2000} (right axis) vs. neutron energy.}
\label{Nspec}
\end{figure}

In order to compare this measured flux with theoretical predictions, the neutron flux expected at the neutron detector from a beam halo of power $P_H$ hitting a simplified aperture block placed inside a thick steel beam pipe was modeled using the MCNP code~\cite{MCNP}. The resulting neutron flux density was
\begin{equation}
\frac{dN_n}{dA\cdot dt}=3300\frac{\rm neutrons}{cm^2Ws}\cdot P_H
\end{equation}
\noindent
The corresponding neutron energy spectrum was folded numerically with the response function (effective dose conversion factor) $c_n$ of the neutron detector~\cite{Pell2000}, both shown in Fig.~\ref{Nspec} which yielded an average value for $c_n$ of about 320 pSv cm$^2$. Multiplying this value with the expected neutron flux density of eqn. (5), we obtained an expected dose rate of
\begin{equation}
R_n\approx0.38[\frac{\rm rem}{(Wh)}]\cdot P_H
\end{equation}
\noindent
which is about 15\% below the measured value of eqn. (4).

Applying the same relation $R_n(P_B,P_W)$ of eqn. (4) to runs 1 and 2 for 6~mm and 4~mm apertures, the resulting wake field powers $P_W$  deposited are 0.1 W for the 4~mm and 0.08 W for the 6~mm aperture.

Table~\ref{Target_comp} shows a summary of average results for the four transmission runs.

\begin{table}[htbp]
\centering
\begin{tabular}{|c|c|c|c|c|c|c|}
\hline\hline
Run & apert. & $P_B(W)$ & $P_W(W)$ & $P_H(W)$ & $I_{ave}(mA)$ & beam loss \\
\hline
        &  mm    &   W    &  W   &  W   &  mA   & ppm\\
\hline\hline
1 & 6  & 0.33 & 0.08 & 0.5 & 3.94 & 1.3\\
2 & 4  & 0.57 & 0.10 & 0.9 & 3.87 & 2.4\\
3 & 2  & 1.95 & 0.50 & 2.9 & 4.25 & 6.8\\
4 & 2  & 1.09 & 0.50 & 1.2 & 4.23 & 2.8 \\
\hline\hline
\end{tabular}
\caption{Wake field power and beam transmission losses}
\label{Target_comp}
\end{table}
In summary, the measurements reported here indicate that a 100~MeV electron beam of 0.43 MW average power can be passed indefinitely  through a 2~mm diameter aperture of 127 mm length with an average beam loss of about 3 ppm with an estimated uncertainty of $\pm20$ percent.  This level of losses is acceptable for the DarkLight experiment and the beam backgrounds generated are manageable.  Substantial improvement on this performance can be achieved by employing both movable and fixed collimators upstream of the experiment, as has been established in experiments with internal gas targets in electron storage rings.  In addition, further optimization of the tune of the beam can be expected.  Thus, luminosities of order 10$^{36}$ nucleons  cm$^{-2}$ s$^{-1}$ are expected to be achievable with Megawatt electron beams of energy $\sim$ 100 MeV from ERLs incident on windowless gas targets. 

We gratefully acknowledge the outstanding efforts of both the staff of the Jefferson Laboratory to deliver the high quality FEL beam and the staff of the MIT-Bates Research and Engineering Center who designed, constructed and delivered the test target assembly.  The research is supported by the United States Department of Energy Office of Science.

\end{document}